\documentclass[aps,prb,reprint,amsfonts,amsmath,amssymb,showpacs,groupedaddress,superscriptaddress]{revtex4-2}
\usepackage[colorlinks,urlcolor=blue,linkcolor=blue,anchorcolor=blue,citecolor=blue,bookmarks]{hyperref}
\usepackage{graphicx}
\usepackage{bm}
\usepackage{color}
\usepackage{float}
\usepackage{subfigure}

\usepackage{xcolor}
\usepackage{ulem}
\usepackage{soul}

\begin{document}

\title {Nonreciprocal impurity scattering as a probe for pairing symmetries in kagome superconductors}

\author{Hong-Min Jiang}
\email{hmjiang@zust.edu.cn} \affiliation{Department of Physics, Zhejiang
University of Science and Technology, Hangzhou 310023, China}
\author{Hao Du}
\affiliation{Department of Physics, Zhejiang University of Science and
Technology, Hangzhou 310023, China}
\author{Shun-Li Yu}
\email{slyu@nju.edu.cn} \affiliation{National Laboratory of Solid
State Microstructures and Department of Physics, Nanjing University,
Nanjing 210093, China} \affiliation{Collaborative Innovation Center
of Advanced Microstructures, Nanjing University, Nanjing 210093,
China} \affiliation{Jiangsu Key Laboratory of Quantum Information Science and Technology, Nanjing University, Suzhou 215163, China}

\date{\today}

\begin{abstract}
The superconducting (SC) pairing symmetry and its link to time-reversal symmetry breaking (TRSB) in the vanadium-based kagome superconductors remain unresolved, with ambiguities stemming from sublattice interference and charge-density-wave (CDW) entanglement with superconductivity.
Using two representative SC pairings, i.e., the conventional on-site $s$-wave and the TRSB $d_{x^2-y^2}+id_{xy}$-wave, as a model study, we theoretically show that while single magnetic impurity yield qualitatively identical spectral behavior of local density of states (LDOS) for these two symmetries, two magnetic impurities give rise to distinct LDOS patterns. For the conventional on-site $s$-wave pairing, time-reversal symmetry (TRS) enforces equivalent forward and backward scattering between two impurities across all impurity configurations, leading to near disappearance of a Yu-Shiba-Rusinov (YSR) state pair along the line connecting the two impurities. However, for the TRSB $d_{x^2-y^2}+id_{xy}$-wave pairing, this scattering equivalence holds only for inversion-symmetric impurity configurations, with a pair of YSR disappearance restricted to this case. These distinct spectral features are resolvable in scanning tunneling microscopy (STM) experiments, providing a direct avenue to discriminate TRSB and non-TRSB SC pairing symmetries in kagome superconductors and an alternative method to probe SC nonreciprocity that circumvents the ambiguities of conventional critical current-based techniques.
\end{abstract}

\maketitle

\section{\label{intro}Introduction}

Kagome systems, characterized by their frustrated lattice geometry, nontrivial band topology and unique sublattice correlation, have long served as paradigmatic platforms for investigating rich quantum states of electronic matter, including spin liquid~\cite{PhysRevLett.98.117205,PhysRevLett.101.117203,doi:10.1126/science.1201080,
PhysRevLett.109.067201,Han2012,PhysRevLett.112.137202,PhysRevLett.118.137202,Feng_2017,Khuntia2020}, superconductivity~\cite{PhysRevB.79.214502,PhysRevB.85.144402,PhysRevB.86.121105,PhysRevLett.110.126405,
PhysRevB.87.115135}, various topological quantum phases~\cite{PhysRevB.80.113102,PhysRevB.80.193304,PhysRevB.82.075125,PhysRevLett.106.236802,
PhysRevLett.115.186802,Ye2018,doi:10.1126/science.aav2873,Yin2020}, charge density wave (CDW)~\cite{PhysRevB.80.113102,PhysRevB.83.165118}, spin-density wave~\cite{PhysRevB.80.193304},
and bond density wave~\cite{PhysRevLett.110.126405,PhysRevB.87.115135,PhysRevLett.97.147202}. The recent discovery of series
of compounds AV$_{3}$Sb$_{5}$ (A=K, Rb or Cs), which share a
common lattice structure with kagome net of vanadium atoms, exhibit a cascade of symmetry-breaking transitions~\cite{Zhao2021Cascade}, involving the $3Q$ chiral charge ordering~\cite{PhysRevX.11.031050,PhysRevX.11.041030,Mielke2022,
Jiang2021Unconventional,Guo2022Switchable}, electronic nematicity~\cite{Nie2022Charge,Li2022Rotation}, roton pair density wave~\cite{Chen2021Roton} and superconductivity~\cite{PhysRevLett.125.247002}.

Of particular interest is the superconducting (SC) state that emerges near van Hove filling (VHF). Driven by intrinsic sublattice textures, this SC state can overcome both lattice frustration and the tendency toward density-wave instabilities, even with a perfectly nested Fermi surface (FS)~\cite{PhysRevB.85.144402}. Despite extensive studies on this system, the microscopic nature of the SC ground state in such a kagome material remains an outstanding open question. Crucially, a consensus on the electron pairing symmetry, essential for understanding the SC state's character, has not yet been reached.

Although most experimental evidence on the AV$_{3}$Sb$_{5}$ family of materials points to a conventional origin with spin-singlet $s$-wave superconductivity~\cite{CMu1,WDuan1,YZhong1,YXie1},
the field remains divided by seemingly conflicting experimental reports on the pairing symmetry. Several experiments reveal an anisotropic SC gap rather than an isotropic one, hinting to an unconventional SC pairing mechanism. Furthermore, signatures of time-reversal symmetry breaking (TRSB) observed in these materials~\cite{Mielke2022,Wu2022,Guguchia2023,doi:10.1126/sciadv.adg7269,TLe1} persist into the SC state. Notably, this occurs even when charge-density-wave (CDW) order is fully suppressed by Ta substitution in CsV$_{3}$Sb$_{5}$~\cite{HDeng1}, suggesting a direct link between TRSB and the SC state~\cite{doi:10.1126/sciadv.adg7269,TLe1}.

The current challenge arises, on the one hand, from the sublattice interference effect in kagome superconductors, which causes all conceivable
unconventional spin-singlet pairings to exhibit excitation characteristics qualitatively similar to the conventional $s$-wave superconductors~\cite{PhysRevB.106.014501,PhysRevB.108.144508,PhysRevB.110.144516,74gh-tjmq}. On the other hand, the CDW order may become entangled with the SC order, leading to considerable debate over whether the signatures of broken time-reversal symmetry (TRS) observed in the SC state originate intrinsically from the SC state itself or are instead a legacy of the CDW order.

In this paper, we investigate impurity-induced in-gap states and the interference between two parallel-aligned magnetic impurities in kagome superconductors, focusing on two representative SC pairing symmetries. One corresponds to the conventional on-site $s$-wave pairing, while the other is a TRSB pairing, namely the $d_{x^{2}-y^{2}}+id_{xy}$-wave~\cite{PhysRevB.108.144508} or equivalently the frustrated SC pairing introduced in Ref.~\cite{74gh-tjmq}. Our calculations reveal that single magnetic impurities yield qualitatively identical impurity-induced behavior for these two distinct SC pairings, whereas two magnetic impurities give rise to markedly different impurity-induced spectral patterns for the two symmetries. For on-site conventional $s$-wave pairing, TRS enforces equivalence between the scattering process from impurity $1$ to impurity $2$ and its time-reversed counterpart. This symmetry constraint results in the near disappearance of a pair of Yu-Shiba-Rusinov (YSR) states along the line connecting the two impurities for all impurity configurations considered. In contrast, the $d_{x^{2}-y^{2}}+id_{xy}$-wave pairing breaks TRS, such that the aforementioned forward and reversed scattering processes are equivalent only for impurity configurations related by inversion symmetry about the impurities’ midpoint. Correspondingly, the near disappearance of the YSR state's pair is observed exclusively in this symmetric configuration. These distinct spectral signatures offer a direct means to distinguish SC pairing symmetries with and without TRSB. Beyond this, they establish an alternative approach to detect nonreciprocal behavior in superconductors, thereby overcoming a critical limitation of conventional characterization techniques for probing TRSB and nonreciprocity in SC systems.

The remainder of the paper is organized as follows. In Sec. II, we
introduce the model Hamiltonian and carry out analytical
calculations. In Sec. III, we present numerical calculations and
discuss the results. In Sec. IV, we make a conclusion.

\section{model and method}
\begin{figure}
\begin{center}
\vspace{.2cm}
\includegraphics[width=230pt,height=210pt]{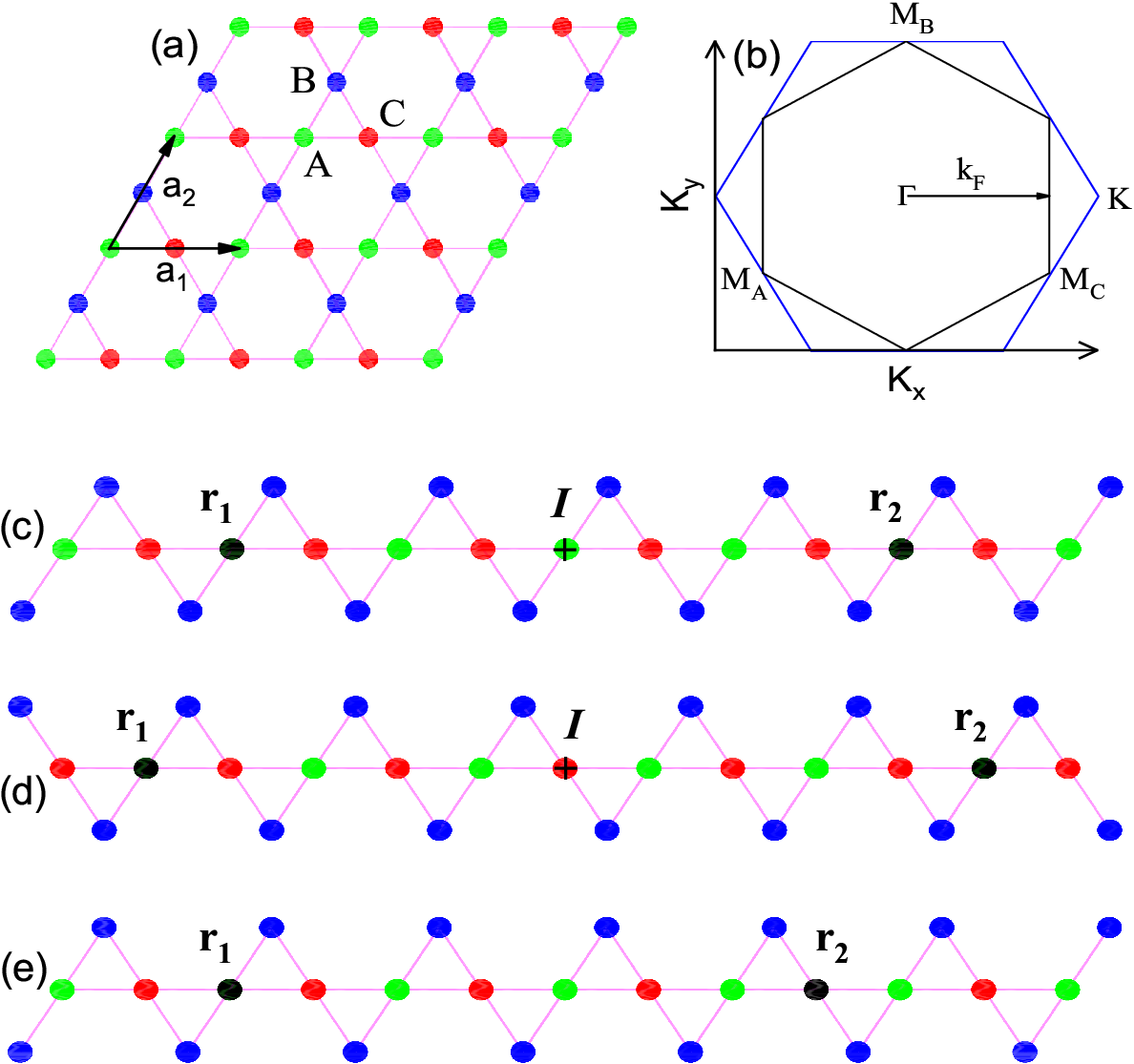}
\caption{(a) Structure of the kagome lattice with lattice vectors $\mathbf{a}_{1}$ and $\mathbf{a}_{2}$, composed of three sublattices denoted by $A$ (green dots), $B$ (red dots), and $C$ (blue dots). (b) Fermi surface generated by the Hamiltonian at a doping level of $1/6$ (black solid lines), with the Fermi wave vector satisfying $k_{F}=\pi$ along the $x$-axis. Impurity configuration with two magnetic impurities (black dots) located on equivalent $A$-sublattice sites at a separation of $R_{i}=4|\mathbf{a}_{1}|$ (c), and $R_{i}=5|\mathbf{a}_{1}|$ (d), respectively. The black cross marked $I$ in panels (c) and (d) denotes the inversion center of the two impurities. (e) Impurity configuration with two magnetic impurities (black dots) located on distinct $A$- and $C$-sublattice sites at a separation of $R_{i}=3.5|\mathbf{a}_{1}|$.}
\label{fig1}
\end{center}
\end{figure}

We consider a kagome model Hamiltonian that is generally used in
the previously investigations on the kagome superconductors. The
Hamiltonian in the normal state reads,
\begin{eqnarray}
H_{0}(\mathbf{k})&=&\sum_{\mathbf{k}\alpha}\hat{\Phi}^{\dag}_{\mathbf{k}\alpha}\hat{\mathcal{H}}^{0}_{\mathbf{k}}
\hat{\Phi}_{\mathbf{k}\alpha},
\end{eqnarray}
with
$\hat{\Phi}_{\mathbf{k}\alpha}=(c_{A\mathbf{k}\alpha},c_{B\mathbf{k}\alpha},c_{C\mathbf{k}\alpha})^{T}$
and
\begin{eqnarray}
\hat{\mathcal{H}}^{0}_{\mathbf{k}}=\left(
\begin{array}{ccc}
-\mu & -2t\cos k_{1} &
-2t\cos k_{2} \\
-2t\cos k_{1} & -\mu & -2t\cos k_{3} \\
-2t\cos k_{2} & -2t\cos k_{3} & -\mu
\end{array}
\right).
\end{eqnarray}
The subscript $m=A,B,C$ in $c_{mk\alpha}$ labels the three basis sites
in the triangular primitive unit cell as shown in
Fig.~\ref{fig1}(a), and $\alpha$ denotes the spin index. $k_{n}$ is abbreviated from
$\mathbf{k}\cdot\mathbf{\tau}_{n}$ with
$\mathbf{\tau}_{1}=\mathbf{a}_{1}/2$,
$\mathbf{\tau}_{2}=\mathbf{a}_{2}/2$ and
$\mathbf{\tau}_{3}=\mathbf{\tau}_{2}-\mathbf{\tau}_{1}$ denoting the
three NN vectors. Here, the lattice vectors are defined as
\begin{eqnarray}
\mathbf{a}_{1}=(1,0), \; \; \; \mathbf{a}_{2}=(\frac{1}{2},\frac{\sqrt{3}}{2}).
\end{eqnarray}

Superconductivity is included at the mean-field level
through the Bogoliubov-de Gennes (BdG) framework. In
Nambu basis, the Hamiltonian reads
\begin{eqnarray}
H(\mathbf{k})&=&\sum_{\mathbf{k}\alpha}\hat{\Psi}^{\dag}_{\mathbf{k}\alpha}\hat{\mathcal{H}}_{\mathbf{k}}
\hat{\Psi}_{\mathbf{k}\alpha},
\end{eqnarray}
with
$\hat{\Psi}_{\mathbf{k}\alpha}=(\hat{\Phi}_{\mathbf{k}\alpha},\hat{\Phi}^{\dag}_{-\mathbf{k}\bar{\alpha}})^{T}$
and
\begin{eqnarray}
\hat{\mathcal{H}}_{\mathbf{k}}=\left(
\begin{array}{cc}
\hat{\mathcal{H}}^{0}_{\mathbf{k}} &
\hat{\Delta}_{\mathbf{k}} \\
\hat{\Delta}^{\dag}_{\mathbf{k}} & -\hat{\mathcal{H}}^{0}_{-\mathbf{k}}
\end{array}
\right).
\end{eqnarray}
Here, the SC order parameters $\hat{\Delta}_{\mathbf{k}}$, defined as $3\times 3$ matrices in the orbital basis, take the form
\begin{eqnarray}
\hat{\Delta}_{\mathbf{k}}=\frac{1}{\sqrt{3}}\left(
\begin{array}{ccc}
\Delta & 0 & 0 \\
0 & \Delta & 0 \\
0 & 0 & \Delta
\end{array}
\right)
\end{eqnarray}
for the conventional on-site $s$-wave pairing, and
\begin{eqnarray}
\hat{\Delta}_{\mathbf{k}}=\frac{1}{\sqrt{6}}\left(
\begin{array}{ccc}
-\Delta & 0 & 0 \\
0 & 2\Delta & 0 \\
0 & 0 & -\Delta
\end{array}
\right)+i\frac{1}{\sqrt{2}}\left(
\begin{array}{ccc}
-\Delta & 0 & 0 \\
0 & 0 & 0 \\
0 & 0 & \Delta
\end{array}
\right)
\end{eqnarray}
for the $d_{x^{2}-y^{2}}+id_{xy}$-wave pairing state. In the following calculations, we choose $\Delta=0.2$ as previously done~\cite{PhysRevB.108.144508}.

An impurity with potential strength $U$ and spin $S$ whose interaction with
electrons is described by the Hamiltonian
$H_{im}=\sum_{\mathbf{k}}\Psi^{\dag}_{\mathbf{k}\alpha}\hat{V}_{\alpha\beta}\hat{\Psi}_{\mathbf{k}\beta}$ with $\hat{V}=\hat{\tau}_{3}(U\hat{\sigma}_{0}-V\hat{\sigma})$. Here, the impurity scatting is assumed to be purely local and isotropic with $V=JS$ ($U$) denoting
the magnetic (potential) scattering strength. $\hat{\tau}_{3}$ and $\hat{\sigma}$ are the Pauli
matrices in Nambu and spin space, respectively. To describe the
effects of impurities, we employ the $\hat{T}$-matrix
formalism~\cite{YLu1,Shiba1,Rusinov1}, which treats the impurity spins as classical,
static variables, corresponding to the limit $JS=$const and
$S\rightarrow\infty$.

Within the $\hat{T}$-matrix approach, one has for the full Greens
function
\begin{eqnarray}
\hat{G}(\mathbf{r},\mathbf{r}',i\omega)=&&\hat{G}_{0}(\mathbf{r},\mathbf{r}',i\omega)
+\hat{G}_{0}(\mathbf{r},\mathbf{r}_{i},i\omega)
\nonumber \\
&&\times\hat{T}^{(0)}(\mathbf{r}_{i},\mathbf{r}_{i},i\omega)\hat{G}_{0}(\mathbf{r}_{i},\mathbf{r}',i\omega),
\end{eqnarray}
where the unperturbed Green's function $\hat{G}_{0}(\mathbf{r},\mathbf{r}',i\omega)=\hat{G}_{0}(\mathbf{r}-\mathbf{r}',i\omega)$ is given by
\begin{eqnarray}
\hat{G}_{0}(\mathbf{r}-\mathbf{r}',i\omega)=
\sum_{\mathbf{k}}\hat{G}_{0}(\mathbf{k},i\omega)\exp[i\mathbf{k}\cdot(\mathbf{r}-\mathbf{r}')]
\end{eqnarray}
with $\hat{G}_{0}(\mathbf{k},i\omega)=[i\omega\hat{1}-\hat{\mathcal{H}}_{\mathbf{k}}]^{-1}$. The one-impurity $T$-matrix
$\hat{T}^{(0)}(\mathbf{r}_{i},\mathbf{r}_{i},i\omega)$ is defined as
$\hat{T}^{(0)}(\mathbf{r}_{i},\mathbf{r}_{i},i\omega)\equiv
[\hat{1}-\hat{V}\hat{G}_{0}(\mathbf{r}_{i}-\mathbf{r}_{i},i\omega)]^{-1}\hat{V}$.

When there are two impurities in the system, the $\hat{T}$-matrix
formalism is generalized to two impurities and the full Greens
function can be expressed as,
\begin{eqnarray}
\hat{G}(\mathbf{r},\mathbf{r}',i\omega)=&&\hat{G}_{0}(\mathbf{r},\mathbf{r}',i\omega)
+\sum_{i,j=1}^{2}\hat{G}_{0}(\mathbf{r},\mathbf{r}_{i},i\omega)
\nonumber \\
&&\times\hat{T}(\mathbf{r}_{i},\mathbf{r}_{j},i\omega)\hat{G}_{0}(\mathbf{r}_{j},\mathbf{r}',i\omega),
\end{eqnarray}
where $\mathbf{r}_{1,2}$ are the positions of the impurities, and
$\hat{T}$-
matrix obeys the Bethe-Salpeter equation,
\begin{eqnarray}
\hat{T}(\mathbf{r}_{i},\mathbf{r}_{j},i\omega)=&&\hat{V}\delta_{r_{i},r_{j}}+\sum_{l=1}^{2}\hat{V}
\hat{G}_{0}(\mathbf{r}_{i},\mathbf{r}_{l},i\omega) \nonumber \\
&&\times\hat{T}(\mathbf{r}_{l},\mathbf{r}_{j},i\omega).
\end{eqnarray}
In the presence of two $\delta$-function impurities localized at $\mathbf{r}=\mathbf{r}_{i}$,
$i\in{1,2}$ with respective strengths $V_{1} (U_{1})$ and $V_{2} (U_{2})$, the matrix $\hat{V}$ takes the following form,
\begin{eqnarray}
\hat{V}(\mathbf{r})=\hat{V}_{1}\delta_{r,r_{1}}+\hat{V}_{2}\delta_{r,r_{2}},
\end{eqnarray}
with $\hat{V}_{1}=\hat{\tau}_{3}(U_{1}\hat{\sigma}_{0}-V_{1}\hat{\sigma})$ and $\hat{V}_{2}=\hat{\tau}_{3}(U_{2}\hat{\sigma}_{0}-V_{2}\hat{\sigma})$.

For brevity we omit energy dependence in the following process.
Solving the system of four equations in Eq. (11), we obtain,
\begin{eqnarray}
\hat{T}(11)&=&[\hat{1}-\hat{T}_{1}^{(0)}(11)\hat{G}_{0}(12)\hat{T}_{2}^{(0)}(22)\hat{G}_{0}(21)]^{-1}\hat{T}_{1}^{(0)}(11), \nonumber \\
\hat{T}(12)&=&\hat{T}_{1}^{(0)}(11)\hat{G}_{0}(12)\hat{T}(22), \nonumber \\
\hat{T}(21)&=&\hat{T}_{2}^{(0)}(22)\hat{G}_{0}(21)\hat{T}(11), \nonumber \\
\hat{T}(22)&=&[\hat{1}-\hat{T}_{2}^{(0)}(22)\hat{G}_{0}(21)\hat{T}_{1}^{(0)}(11)\hat{G}_{0}(12)]^{-1}\hat{T}_{2}^{(0)}(22),
\end{eqnarray}
where $\hat{T}(ij)$, $\hat{G}_{0}(ij)$ and $\hat{T}_{i}^{(0)}(ii)$
are the abbreviations for
$\hat{T}(\mathbf{r}_{i},\mathbf{r}_{j},i\omega)$,
$\hat{G}_{0}(\mathbf{r}_{i},\mathbf{r}_{j},i\omega)$ and
$\hat{T}_{i}^{(0)}(\mathbf{r}_{i},\mathbf{r}_{i},i\omega)$, respectively.
After obtaining the full Green's functions, the spin-summed
electronic LDOS $\rho_{\alpha}(\mathbf{r},\omega)$ at sublattice
position $\alpha$ is obtained by
$\rho_{\alpha}(\mathbf{r},\omega)=-\frac{1}{\pi}\textmd{Im}
[\hat{G}_{\alpha\alpha}(\mathbf{r},\mathbf{r},\omega+i\delta)
+\hat{G}_{\bar{\alpha}\bar{\alpha}}(\mathbf{r},\mathbf{r},-\omega+i\delta)]$,
where $\bar{\alpha}$ refers to the same sublattice site as $\alpha$
but in the complementary Nambu block. As the absence of impurity-induced in-gap states from nonmagnetic impurity for all possible even-parity SC pairing states is a generic feature of the kagome superconductors~\cite{PhysRevB.108.144508}, we restrict our subsequent calculations to the case of magnetic impurities.

\section{Results and discussion}

%%%%%%%%%%%%%%%%%%%%%%%%%%%%%%%%%%%%%%%
\vspace*{.2cm}
%%%%%%%%%%%%%%%%%%%%%%%%%%%%%%%%%%%%

\subsection{Single magnetic impurity}
Prior to examining the two-impurity scenario in kagome superconductors, it is instructive to first recap the nature of impurity states across different SC pairing symmetries. A cornerstone of this discussion is Anderson’s theorem for conventional $s$-wave superconductors, which posits that isotropic superconductivity is resistant to pair-breaking induced by nonmagnetic impurities, thus precluding the formation of in-gap states~\cite{ANDERSON195926}. This stands in stark contrast to the effect of magnetic impurities, which can stabilize in-gap bound states known as Yu-Shiba-Rusinov (YSR) states~\cite{YLu1,Shiba1,Rusinov1}. For a single magnetic impurity, the exchange interaction ($V$) of magnetic adatoms with the conduction electrons of the underlying superconductor gives rise to particle- and hole-like excitations within the SC gap.

\begin{figure}
\centering
\includegraphics[width=240pt,height=180pt]{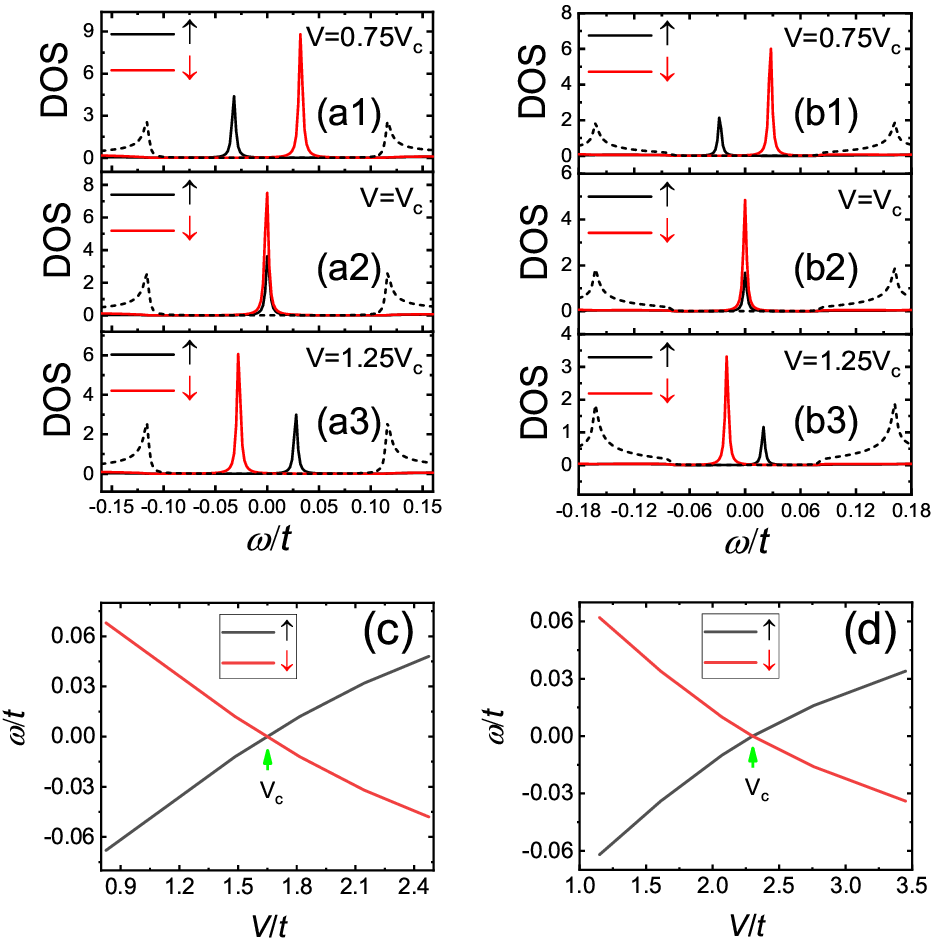}
\caption{Solid curves in panels (a) and (b) show the LDOS at the impurity site for the single-impurity case. The DOSs for the clean case are also shown as dotted curves in each panels. The left column [panels (a1)-(a3)] corresponds to conventional on-site $s$-wave superconductivity, while the left column [panels (b1)-(b3)] illustrates the $d_{x^{2}-y^{2}}+id_{xy}$-wave pairing symmetry. (a1)/(b1), (a2)/(b2), and (a3)/(b3) present the results for $V=0.75V_{c}$, $V=V_{c}$, and $V=1.25V_{c}$, respectively. Here, $V_{c}$ denotes the critical exchange interaction at which particle- and hole-like excitations meet at zero energy, as evidenced in panels (c) and (d). Panels (c) and (d) illustrate the energy evolution of the YSR states as a function of the exchange interaction $V$ for the conventional on-site $s$-wave and the $d_{x^{2}-y^{2}}+id_{xy}$-wave pairings, respectively.}
\label{fig2}
\end{figure}

As illustrated in Fig.~\ref{fig2}, the density of states (DOS) exhibits a symmetric distribution of two peaks around zero energy, where the left and right columns depict two representative SC pairing symmetries, i.e., the conventional on-site $s$-wave and the $d_{x^{2}-y^{2}}+id_{xy}$-wave pairings, respectively.
Notably, a single magnetic impurity exerts nearly identical effects on both pairing symmetries. Specifically, as the coupling strength $V$ increases from zero, the in-gap states shift from the gap edges toward the gap center and cross zero energy at a critical value $V=V_{c}$, thereby triggering a phase transition in which the spin polarization of the superconductor’s ground state
changes~\cite{10.1143/PTP.44.1472,PhysRevB.67.020502,PhysRevB.55.12648,10.1063/1.373404}. The evolution of the peak positions with $V$ for the two pairing symmetries is summarized in Figs.~\ref{fig2}(c) and (d).
In the case of single magnetic impurity, despite the distinct SC pairing symmetries of the two systems, their impurity-induced behaviors are qualitatively identical, rendering them difficult to distinguish.

\subsection{Two magnetic impurities}

To explore the unconventional interference between impurities and enable discrimination of the two most plausible SC pairing symmetries in kagome superconductors, we extend our investigation to the case of two parallel-aligned magnetic impurities, aiming to clarify whether this indistinguishability persists. In the seminal work by Rusinov~\cite{Rusinov2}, it was predicted that Yu-Shiba-Rusinov (YSR) states arising from two nearby parallel-aligned magnetic impurities located on $\mathbf{r}_{1}$ and $\mathbf{r}_{2}$ can hybridize, forming symmetric and antisymmetric combinations of YSR wavefunctions for both the particle-like states $|p,\downarrow\rangle_{s,a}=(|p,\downarrow,\mathbf{r}_{1}\rangle\pm |p,\downarrow,\mathbf{r}_{2}\rangle)$ and the hole-like states $|h,\uparrow\rangle_{s,a}=(|h,\uparrow,\mathbf{r}_{1}\rangle\pm |h,\uparrow,\mathbf{r}_{2}\rangle)$. These symmetric and antisymmetric YSR states exhibit opposite energy shifts, manifesting as a spectroscopic signature characterized by split peaks for both particle-like [denoted as $p_{1}$ and $p_{2}$ in Fig.~\ref{fig3}(c) and (d)] and hole-like [denoted as $h_{1}$ and $h_{2}$ in Fig.~\ref{fig3}(c) and (d)] excitations in the local density of states (LDOS).

For a concrete grasp of impurity interference, we focus on impurity-induced excitation changes through the definition of
\begin{eqnarray}
\delta\hat{G}(\mathbf{r},\mathbf{r},i\omega)\equiv&&\hat{G}(\mathbf{r},\mathbf{r},i\omega)-\hat{G}_{0}(\mathbf{r},\mathbf{r}',i\omega)
\nonumber \\
=&&\sum_{i,j=1}^{2}\hat{G}_{0}(\mathbf{r},\mathbf{r}_{i},i\omega)
\hat{T}(\mathbf{r}_{i},\mathbf{r}_{j},i\omega)\hat{G}_{0}(\mathbf{r}_{j},\mathbf{r},i\omega).
\end{eqnarray}
The right-hand side of Eq. (14) can be divided into two distinct categories. The first category includes all processes that start at $\mathbf{r}$, propagate to $\mathbf{r}_{1}$ ($\mathbf{r}_{2}$), and return to $\mathbf{r}$ from $\mathbf{r}_{1}$ ($\mathbf{r}_{2}$) following potential multiple bounces. Referred to as ``single contributions"~\cite{PhysRevLett.130.256001}, this category yields the impurity-induced changes in the Green's function,
\begin{eqnarray}
\delta\hat{G}_{11(22)}(\mathbf{r},\mathbf{r},i\omega)=&&\hat{G}_{0}(\mathbf{r},\mathbf{r}_{1(2)},i\omega)
\hat{T}(\mathbf{r}_{1(2)},\mathbf{r}_{1(2)},i\omega)
\nonumber \\ &&\times\hat{G}_{0}(\mathbf{r}_{1(2)},\mathbf{r},i\omega),
\end{eqnarray}
and the corresponding variations in the LDOS,
\begin{eqnarray}
\delta\rho_{11(22)}=-\frac{1}{\pi}\textmd{Im}
\{[\delta\hat{G}_{11(22)}(\mathbf{r},\mathbf{r},\omega+i\delta)]_{\alpha\alpha}
\nonumber \\+[\delta\hat{G}_{11(22)}(\mathbf{r},\mathbf{r},-\omega+i\delta)]_{\bar{\alpha}\bar{\alpha}}\}.
\end{eqnarray}
The second category consists of the so-called ``loop contributions"~\cite{PhysRevLett.130.256001},  which start at $\mathbf{r}$, propagate to $\mathbf{r}_{1}$ (or $\mathbf{r}_{2}$), and return to $\mathbf{r}$ from $\mathbf{r}_{2}$ (or $\mathbf{r}_{1}$) after possible multiple bounces,
\begin{eqnarray}
\delta\hat{G}_{21(12)}(\mathbf{r},\mathbf{r},i\omega)=&&\hat{G}_{0}(\mathbf{r},\mathbf{r}_{2(1)},i\omega)
\hat{T}(\mathbf{r}_{2(1)},\mathbf{r}_{1(2)},i\omega)\nonumber \\ &&\times\hat{G}_{0}(\mathbf{r}_{1(2)},\mathbf{r},i\omega).
\end{eqnarray}
with the corresponding variations in the LDOS given by,
\begin{eqnarray}
\delta\rho_{21(12)}=-\frac{1}{\pi}\textmd{Im}
\{[\delta\hat{G}_{21(12)}(\mathbf{r},\mathbf{r},\omega+i\delta)]_{\alpha\alpha}
\nonumber \\+[\delta\hat{G}_{21(12)}(\mathbf{r},\mathbf{r},-\omega+i\delta)]_{\bar{\alpha}\bar{\alpha}}\}.
\end{eqnarray}

Notably, the kagome lattice comprises three inequivalent sublattice sites, denoted as $A$, $B$, and $C$ in Fig.~\ref{fig1}(a). Consequently, two distinct configurations exist for the two impurities: one involves both of them residing on equivalent sublattice sites (i.e., both on $A$, both on $B$, or both on $C$) as illustrated in Fig.~\ref{fig1}(c) and (d), and the other involves them occupying distinct sublattices (e.g., $A-B$, $A-C$, or $B-C$) as shown in Fig.~\ref{fig1}(e). Without loss of generality, we assume the former case corresponds to both impurities residing on the $A$ sublattice sites along the $\mathbf{a}_{1}$ direction, while the latter case involves the two impurities occupying the distinct $A$- and $C$- sublattices along the $\mathbf{a}_{1}$ direction.

\subsubsection{Inversion-Symmetric Impurities}

When both impurities occupy the same $A$-sublattice sites, they are related by inversion symmetry with respect to their midpoint. This implies that the system remains invariant under the interchange of $\mathbf{r}_1$ and $\mathbf{r}_2$ about this midpoint [see Fig.~\ref{fig1}(c), (d)]. Consequently, the scattering loop $\mathbf{r}\to\mathbf{r}_1\to\mathbf{r}_2\to\mathbf{r}$ is identical to its inversion counterpart $\mathbf{r}\to\mathbf{r}_2\to\mathbf{r}_1\to\mathbf{r}$ when evaluated at the midpoint. Owing to inversion symmetry, the $s$-wave and the $d_{x^{2}-y^{2}}+id_{xy}$-wave pairings yield nearly identical $\omega$-dependent behaviors of $\delta\rho$. Specifically, $\delta\rho_{12}$ and $\delta\rho_{21}$ exhibit identical in-phase enhancement, which interferes with $\delta\rho_{11}$ and $\delta\rho_{22}$ to induce well-defined symmetric and antisymmetric YSR states. Given the qualitative equivalence of results for on-site $s$-wave and the $d_{x^{2}-y^{2}}+id_{xy}$-wave pairings, only the former is presented in this subsection.

\begin{figure}
\centering
\includegraphics[width=240pt,height=180pt]{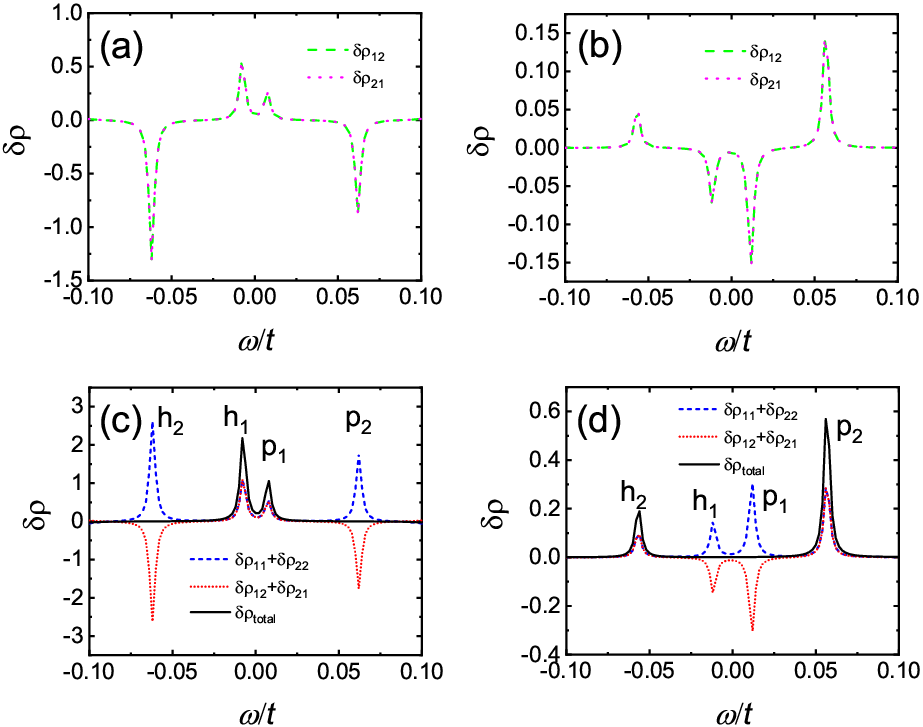}
\caption{Contributions of impurities to the LDOS at the midpoint of the two impurities, arising from $\delta\rho$ for two impurities located on the same $A$ sublattice sites. Panels (a) and (c) show the results at the midpoint site for two impurities situated on the same $A$-sublattice sites with $R_{i}=4|\mathbf{a}_{1}|$. Panels (b) and (d) present the corresponding results for two impurities situated on the same $A$-sublattice sites with $R_{i}=5|\mathbf{a}_{1}|$.}
\label{fig3}
\end{figure}

Two distinct scenarios persist for this configuration. First, the impurity separation corresponds to an even multiple of the lattice constant, i.e., $R_{ie}=|\mathbf{r}_2-\mathbf{r}_1|=2n|\mathbf{a}_{1}|$ ($n\in\mathbb{Z}$) as displayed in Fig.~\ref{fig1}(c). In this case, the inversion center lies on a site of the same $A$ sublattice. Second, the impurity separation equals an odd multiple of the lattice constant, i.e., $R_{io}=(2n+1)|\mathbf{a}_{1}|$ as shown in Fig.~\ref{fig1}(d), where the inversion center is instead located on a site of the distinct $C$ sublattice. Figure~\ref{fig3}(a) and (c) depicts the results for $\delta\rho_{12}$, $\delta\rho_{21}$, $\delta\rho_{11}+\delta\rho_{22}$, $\delta\rho_{12}+\delta\rho_{21}$, and the total impurity-induced perturbation to the local density of states (LDOS) at the midpoint, for on-site $s$-wave pairing with an impurities' separation of $R_{ie}=4|\mathbf{a}_{1}|$. As illustrated in the figure, interference between the ``single contributions" and ``loop contributions" gives rise to constructive interference for the $p_{1}$ and $h_{1}$ excitations (the so-called symmetric YSR states), whereas destructive interference occurs for the $p_{2}$ and $h_{2}$ excitations (the antisymmetric YSR states). Thus, only the $p_{1}$ and $h_{1}$ excitations are detectable in the LDOS at the midpoint, as displayed by solid curve in Fig.~\ref{fig3}(c).

Compared to the even-separation configuration with $R_{ie}=2n|\mathbf{a}_{1}|$, the loop contributions acquire a phase shift of $\Delta\phi=k_{F}(R_{io}-R_{ie})=k_{F}$ when the impurity separation is set to $R_{io}=(2n+1)|\mathbf{a}_{1}|$. At the van Hove filling, the Fermi wavevector along the $x$-axis satisfies $k_{F}=\pi$, as illustrated in Fig.~\ref{fig1}(b). Consequently, the loop contributions for the odd-separation case exhibit a $\pi$ phase shift relative to the even-separation case, as shown in Fig.~\ref{fig3}(b). This results in a complete reversal of the interference pattern for the odd-separation configuration with $R_i=5|\mathbf{a}_{1}|$, where the $p_{1}$ and $h_{1}$ states evolve into antisymmetric YSR states, whereas the $p_{2}$ and $h_{2}$ states become symmetric YSR states, as demonstrated in Fig.~\ref{fig3}(d).

For other interstitial sites along the line connecting the two impurities, displaced by $\Delta\mathbf{r}$ from the midpoint toward $\mathbf{r}_2$, the wave functions $|p,\downarrow,\mathbf{r}_1\rangle$ and $|p,\downarrow,\mathbf{r}_2\rangle$ develop phase shifts of $\Delta\phi_1=k_F\Delta r$ and $\Delta\phi_2=-\Delta\phi_1$, respectively, relative to their midpoint counterparts. For the even-separation configuration, interstitial sites belong to the $A$ sublattice yield $\Delta\phi_1=n\pi$, giving a total phase difference of $2n\pi$ between the two wave functions, and thus leaving them in phase with those at the midpoint. In contrast, interstitial sites on the $C$ sublattice give $\Delta\phi_1=(n+1/2)\pi$, leading to a total phase difference of $(2n+1)\pi$, and thus rendering them out of phase with those at the midpoint.
As a consequence, the $\delta\rho$ spectra at $A$-sublattice interstitial sites reproduce the features shown in Fig.~\ref{fig3}(c), while those at $C$-sublattice sites mirror the behavior displayed in Fig.~\ref{fig3}(d). For the odd-separation configuration, the same arguments apply upon simply interchanging the labels $A$ and $C$ of the interstitial sites.
Therefore, for two impurities located on the same $A$ sublattice, $p_{1}$ and $h_{1}$ excitations always undergo constructive interference, whereas $p_{2}$ and $h_{2}$ excitations experience destructive interference at interstitial $A$-sublattice sites [Fig.~\ref{fig3}(c)]. The interference pattern is inverted at interstitial $C$ sublattice sites [Fig.~\ref{fig3}(d)].

For sites lying on the line outside the range bounded by the two impurities, the phase difference between $|p,\downarrow,\mathbf{r}_1\rangle$ and $|p,\downarrow,\mathbf{r}_2\rangle$ depends solely on the impurity separation, i.e., $\Delta\phi=k_F R_i$. Under this condition, constructive interference for $p_{1}$ and $h_{1}$ excitations and destructive interference for $p_{2}$ and $h_{2}$ excitations are consistently observed at both $A$- and $C$-sublattice sites [Fig.~\ref{fig3}(c)].

\subsubsection{Inversion Symmetry-Breaking Impurities}

\begin{figure}
\centering
\includegraphics[width=240pt,height=180pt]{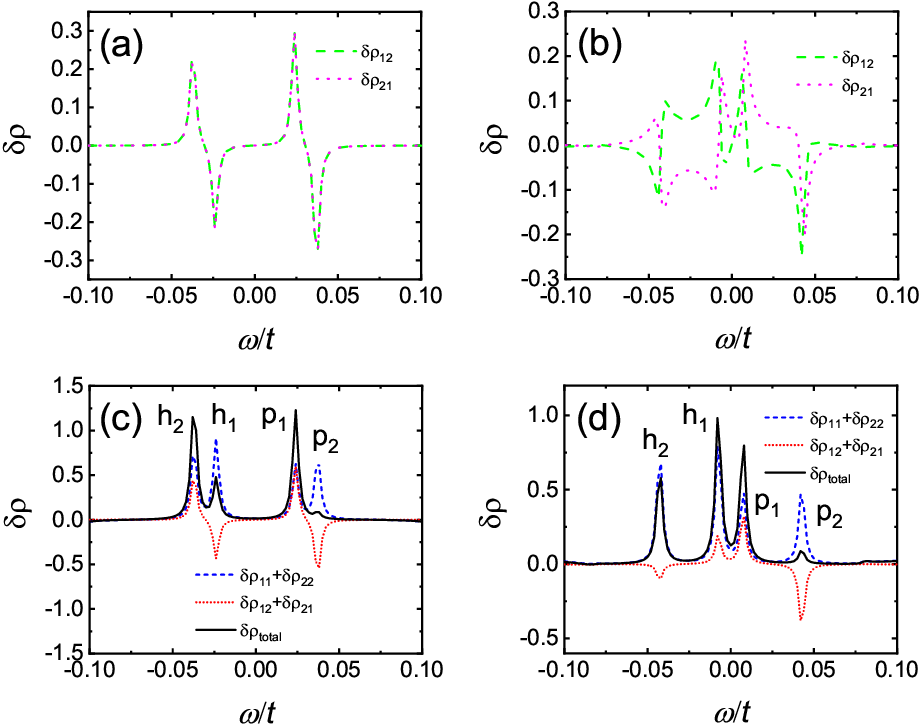}
\caption{Contributions of impurities to the LDOS on interstitial sites between the two impurities, arising from $\delta\rho$ for two impurities located on the distinct $A$- and $C$-
sublattice sites with impurities' separation $R_{i}=3.5|\mathbf{a}_{1}|$. Panels (a) and (c) show the results for the on-site $s$-wave paring. Panels (b) and (d) present the corresponding results for the $d_{x^{2}-y^{2}}+id_{xy}$-wave pairing.}
\label{fig4}
\end{figure}

When two impurities reside on the distinct $A$- and $C$-sublattice sites as presented in Fig.~\ref{fig1}(e), no inversion center is associated with this pair of impurities. Consequently, no well-defined symmetric or antisymmetric YSR states can be formed. Furthermore, since interstitial sites for the $A$ and $C$ sublattices share the same physical environment, the impurity-induced perturbations to the LDOS at these sites are essentially identical. This differs from the inversion-symmetric impurity case, where the ``loop contributions" exhibit opposite phases at the $A$- and $C$-sublattice sites.
For systems preserving TRS, for example in the case of the on-site $s$-wave pairing, the scattering loop $\mathbf{r}\to\mathbf{r}_1\to\mathbf{r}_2\to\mathbf{r}$ and its time-reversed counterpart $\mathbf{r}\to\mathbf{r}_2\to\mathbf{r}_1\to\mathbf{r}$ act as reciprocal processes. Figure~\ref{fig4}(a) shows the results for the on-site $s$-wave pairing with two impurities situated on the distinct $A$- and $C$-sublattice sites with an impurity separation
of $R_{i}=3.5|\mathbf{a}_{1}|$. Here, $\delta\rho_{12}$ and $\delta\rho_{21}$ are found to be identical, yet they differ markedly from the corresponding values for inversion-symmetric impurities [Fig.~\ref{fig3}(a) and (b)]. Specifically, the inner excitations $p_{1}$ and $h_{1}$ arising from the ``loop contributions" exhibit identical phases in the inversion-symmetric case, whereas their phases are opposite for the inversion-symmetry-breaking impurity configuration. This phase difference enhances significantly the inner particle excitation $p_{1}$ and the outer hole excitation $h_{2}$, while causes a nearly complete destructive interference for the outer particle excitation $p_{2}$ and a substantial reduction in the inner hole excitation $h_{1}$, as illustrated in Fig.~\ref{fig4}(c).

Notably, for the $d_{x^{2}-y^{2}}+id_{xy}$-wave pairing, the scattering loop $\mathbf{r}\to\mathbf{r}_1\to\mathbf{r}_2\to\mathbf{r}$ and its time-reversed counterpart display nonreciprocal behavior. This behavior is evident in Fig.~\ref{fig4}(b), where $\delta\rho_{12}$ and $\delta\rho_{21}$ diverge and even exhibit sign-opposing energetic evolution for both particle and hole excitations. The nonreciprocal evolution of $\delta\rho_{12}$ and $\delta\rho_{21}$ results in markedly weak peaks for the ``loop contributions" of the hole excitations $h_{1}$ and $h_{2}$, while partially restoring the phase matching of the inner $p_{1}$ and $h_{1}$ peaks, an effect indicated by the red dotted curve in Fig.~\ref{fig4}(d). Consequently, similar to the on-site $s$-wave pairing case, the $d_{x^{2}-y^{2}}+id_{xy}$-wave pairing exhibits, on the one hand, a substantial enhancement of the inner particle excitation $p_{1}$ and nearly complete destructive interference of the outer particle excitation $p_{2}$, as indicated by the black solid line in Fig.~\ref{fig4}(d). On the other hand, the slight enhancement of $h_{1}$ and minor reduction of $h_{2}$ characteristic of the $d_{x^{2}-y^{2}}+id_{xy}$-wave case differ distinctly from the corresponding behavior observed in the on-site $s$-wave pairing system.

\begin{figure}
\centering
\includegraphics[width=240pt,height=180pt]{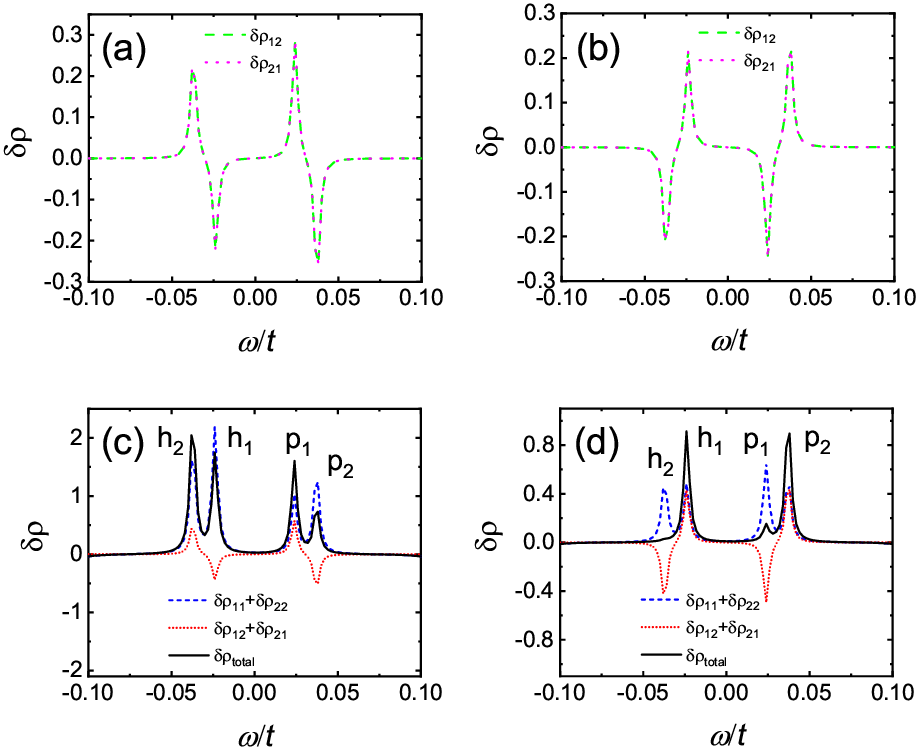}
\caption{Contributions of impurities to the LDOS for the on-site $s$-wave pairing at lattice sites along a line outside the boundary of the two impurities on distinct $A$- and $C$-sublattice sites with $R_{i}=3.5|\mathbf{a}_{1}|$. Panels (a) and (c) show the results at the $C$-sublattice site near the $C$-sublattice impurity. Panels (b) and (d) present the corresponding results at the $A$-sublattice site near the $C$-sublattice impurity.}
\label{fig5}
\end{figure}

\begin{figure}
\centering
\includegraphics[width=240pt,height=180pt]{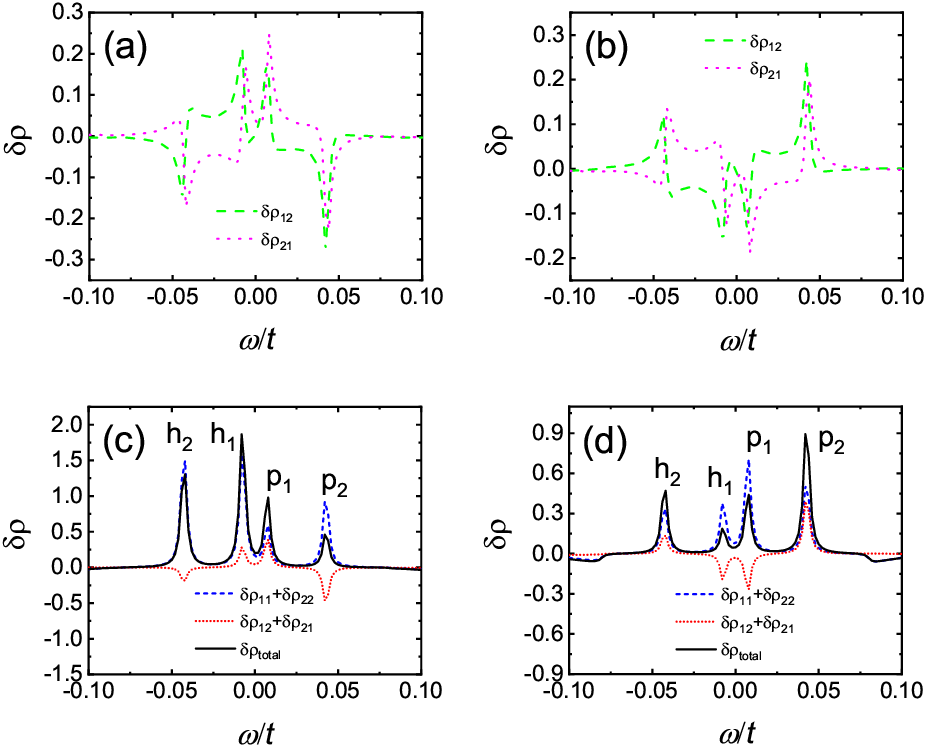}
\caption{Contributions of impurities to the LDOS for the $d_{x^{2}-y^{2}}+id_{xy}$-wave pairing at lattice sites along a line outside the boundary of the two impurities on distinct $A$- and $C$-sublattice sites with $R_{i}=3.5|\mathbf{a}_{1}|$. Panels (a) and (c) show the results at the $C$-sublattice site near the $C$-sublattice impurity. Panels (b) and (d) present the corresponding results at the $A$-sublattice site near the $C$-sublattice impurity.}
\label{fig6}
\end{figure}

More striking distinctions between the two SC pairing symmetries emerge for the impurities' perturbations to the LDOS on the lattice sites along a line outside the boundary demarcated by the two impurities. In this case, the ``loop contributions" ($\delta\rho_{12}$ and $\delta\rho_{21}$) at $A$- and $C$-sublattice sites evolves in opposite fashions for both the on-site $s$-wave and the $d_{x^{2}-y^{2}}+id_{xy}$-wave pairing [Fig.~\ref{fig5}(a), (b) and Fig.~\ref{fig6}(a), (b)], a behavior that also contrasts with the inversion-symmetric case. In Figs.~\ref{fig5} and \ref{fig6}, panels (a) correspond to the $C$-sublattice sites, while panels (b) correspond to the $A$-sublattice sites.

For the on-site $s$-wave pairing, the peaks of the ``single contributions” at $C$-sublattice sites dominate over those of the ``loop contributions” [Fig.~\ref{fig5}(c)], whereas at $A$-sublattice sites both contributions remain comparable [Fig.~\ref{fig5}(d)]. As a result, the full set of four impurity-induced peaks remains intact in the former case, whereas the $p_{1}$ and $h_{2}$ peaks are almost completely suppressed in the latter. By contrast, for the $d_{x^{2}-y^{2}}+id_{xy}$-wave pairing, the ``loop contribution” peaks are several times weaker than the ``single contribution” peaks at both $A$- and $C$-sublattice sites, owing to the nonreciprocal character of $\delta\rho_{12}$ and $\delta\rho_{21}$ [Fig.~\ref{fig6}(a), (b)]. This leads to the full set of four impurity-induced LDOS peaks being clearly resolved at both sublattice sites [Fig.~\ref{fig6}(c), (d)].

On the one hand, the distinct spectral features identified in this work can be directly resolved in scanning tunneling microscopy (STM) experiments and can therefore be utilized to discriminate SC pairing symmetries associated with TRSB. On the other hand, these features provide an alternative avenue to probe nonreciprocal responses in superconductors, a capability that alleviates a critical limitation of conventional characterization techniques for probing TRSB and nonreciprocity in SC systems.

Conventionally, the nonreciprocal behavior in superconductors, known as the SC diode effect, is typically characterized and probed by means of the critical current~\cite{Wu2022,TLe1}. However, other unconventional ordered phases, such as loop-current charge order proposed in the vanadium-based kagome superconductors, can also induce nonreciprocal modulations of the critical current. This ambiguity calls into question the validity of inferring TRSB in the SC state solely from measurements of the SC diode effect in these materials. In contrast, since YSR states arise exclusively from impurity-induced excitations in SC state, the unequal contributions of $\delta\rho_{21}$ and $\delta\rho_{12}$ along with the resulting distinct impurity-perturbed LDOS patterns constitute the most direct and unambiguous signature of TRSB in the SC state.

\section{Conclusion}

In conclusion, we have investigated the impurity-induced in-gap states and the inter-impurity interference between two parallel-aligned magnetic impurities in kagome superconductors for two representative SC pairing symmetries, i.e., the on-site conventional $s$-wave and the $d_{x^2-y^2}+id_{xy}$-wave pairings. Our calculated results showed that while impurity-induced spectral behavior is qualitatively identical for these two distinct pairing symmetries in the case of single magnetic impurity, distinct impurity-induced spectral features emerge in the two magnetic impurity case. For the on-site conventional $s$-wave pairing, TRS enforces equivalence between the scattering process from impurity 1 to impurity 2 and its reverse, leading to the near disappearance of a pair of YSR states along the line connecting the two impurities for all impurity configurations. In contrast, the $d_{x^2-y^2}+id_{xy}$-wave pairing breaks TRS, such that these two scattering processes are equivalent only for impurity configurations related by inversion symmetry about their midpoint with the concomitant near disappearance of the YSR state pair restricted to this specific case. These distinct spectral signatures can be utilized to discriminate SC pairing symmetries associated with TRSB. Additionally, they establish an alternative avenue to probe nonreciprocal responses in superconductors, a capability that mitigates a key limitation of conventional characterization techniques for TRSB and nonreciprocity in SC systems.

\begin{acknowledgements}
This work was supported by the National Key Projects for
Research and Development of China (No. 2024YFA1408104) and the
National Natural Science Foundation of China (No. 12374137, and No.
12434005).
\end{acknowledgements}

\bibliography{ref}

\end{document}